# Graphene Spintronics

Wei Han[1,2,3], Roland K. Kawakami[4,5,†], Martin Gmitra[6], and Jaroslav Fabian[6††]

[1]International Center for Quantum Materials, School of Physics, Peking University, Beijing 100871, P. R. China

[2]Collaborative Innovation Center of Quantum Matter, Beijing 100871, P. R. China

[3] IBM Almaden Research Center, San Jose, California 95120, USA

[4] Department of Physics, The Ohio State University, Columbus, Ohio 43210, USA

[5] Department of Physics and Astronomy, University of California, Riverside, California 92521, USA

[6]Institute for Theoretical Physics, University of Regensburg, D-93040 Regensburg, Germany

[†] e-mail: kawakami.15@osu.edu

[††]e-mail: jaroslav.fabian@ur.de

**The isolation of graphene has triggered an avalanche of studies into the spin-dependent physical properties of this material, as well as graphene-based spintronic devices. Here we review the experimental and theoretical state-of-art concerning spin injection and transport, defect-induced magnetic moments, spin-orbit coupling and spin relaxation in graphene. Future research in graphene spintronics will need to address the development of applications such as spin transistors and spin logic devices, as well as exotic physical properties including topological states and proximity-induced phenomena in graphene and other 2D materials.**

Spintronics aims to utilize the spin degree freedom of electrons for novel information storage and logic devices[1]. Recently, there has been great interest in spin logic devices[2,3] for high speed, low power operation, and spin transistors[4] for reconfigurable logic. For this purpose, a major challenge is developing a suitable spin transport channel with long spin lifetime and long distance spin propagation. Graphene, a single atomic layer of graphitic carbon, is a very promising spin channel material due to the achievement of room temperature spin transport with long spin diffusion lengths of several micrometers[5-10]. Moreover, graphene has many interesting physical properties that also make it very attractive for spintronics, including gate tunable carrier concentration and high electronic mobility[11,12].

In the last decade since the isolation of graphene[13], there have been many significant advances in the field of graphene spintronics, including efficient spin injection into graphene, defect-induced magnetism in graphene, theoretical understanding of the intrinsic and extrinsic spin-orbit coupling, and the investigation of the spin relaxation in graphene. This article focuses on reviewing these advances and is organized as follows. The first section describes the current status of efficient injection of spin polarized carriers into graphene. The second section reviews of magnetic moments in graphene induced by adatoms and vacancy defects. The third section describes both intrinsic and extrinsic spin-orbit couplings in graphene from a theoretical perspective. The fourth section focuses on the investigation of the spin lifetimes and the spin relaxation mechanisms in graphene. The final section discusses the potential applications and future perspectives of graphene spintronics.

## 1. Electrical Spin Injection and Transport in Graphene

To perform electrical spin injection into spin channel materials such as graphene, two different types of measurements known as "nonlocal" and "local" have been commonly used. For the nonlocal measurement (Fig. 1a), a current source is applied between the electrodes, E1 and E2, where E2 serves as a spin injector due to the spin-dependent density of states in a ferromagnetic (FM) metal at the Fermi energy[14,15]. After spin injection, the spins in graphene underneath E2 are able to diffuse in both directions, toward E1 (as a spin current with charge current) and toward E3 (as a spin current without charge current). Then, the spin is detected by measuring the voltage across E3 and E4, where E3 (FM) is the spin detector. This measurement is called "nonlocal" because the voltage probes lie outside the charge current loop; this geometry allows the voltage to detect the spin density at E3 arising from the pure spin current of diffusion of spin-polarized electrons. The measured voltage ($V_{NL}$) is positive or negative depending on whether the magnetization configuration of E2 and E3 is parallel or antiparallel to each other. The difference between these two voltages is the nonlocal spin signal, and it is often converted to units of resistance by dividing out the injection current ($\Delta R_{NL} = (V_{NL}^{P} - V_{NL}^{AP})/I_{inj}$). The nonlocal resistance ($R_{NL} = V_{NL}/I_{inj}$) is not a resistance in the usual sense; it is a four-terminal resistance which can have positive or negative values depending on the polarity of spin density in graphene. The electrodes E1 and E4 are ideally nonmagnetic, but are sometimes FM to simplify the device fabrication process.

The local measurement directly measures the standard two-terminal resistance across two ferromagnetic electrodes (e.g. E2 and E3) as shown in Fig. 1e. Spin polarized electrons are injected from one electrode, transported across the graphene, and detected by the second electrode. The difference of the resistance ($\Delta R$) between the parallel and anti-parallel magnetization alignments of the two electrodes is the local magnetoresistance (MR), which is the signal of spin transport.

It is worth noting that the resistance change in local and nonlocal geometries both arise from spin accumulation and are closely related fundamentally ($\Delta R = 2\ \Delta R_{NL}$)[16], but the nonlocal geometry benefits from higher signal-to-noise due to the absence of net charge flow between injector and detector.

Nonlocal electrical spin transport in graphene at room temperature was first demonstrated in 2007 by the van Wees group (Fig. 1b)[5]. To generate the parallel and antiparallel magnetization alignments of E2 and E3, an in-plane magnetic field is applied along the long axes of the FM electrodes. As the magnetic field is swept up from negative to positive, one electrode, E3, switches the magnetization direction first, resulting in a switching of nonlocal resistance ($R_{NL}=V_{NL}/I_{INJ}$) as the magnetization alignment changes from parallel to anti-parallel. Then, the electrode E2 switches the magnetization direction, and the state changes back to the parallel state. The difference in $R_{NL}$ between the parallel and antiparallel states is called the nonlocal MR (ie. $\Delta R_{NL}$) and is a result of the spin diffusion from E2 to E3. The minor jumps in the nonlocal MR loop are due to the switching of the FM electrodes E1 and E4. A key aspect of this initial study is the observation of Hanle spin precession in the nonlocal geometry (this will be described in a later section, see Fig. 4), which provided unambiguous proof of spin injection and transport in graphene. While other effects could generate voltage changes due to magnetization reversal (e.g. local Hall effect[17]), only spin polarized carriers in the graphene could generate a spin precession signal. The spin polarization of the injected carriers (spin injection efficiency) is estimated to be around 10%.

Subsequent studies[18-24] also exhibited relatively low spin injection efficiency, which could be due to the conductance mismatch problem[25] resulting from high conductance of the FM metal and lower conductance of graphene, or other possible contact-related effects (e.g interfacial spin flip scattering, inhomogeneous dephasing, etc.). High spin injection efficiency was achieved in 2010 by inserting atomically smooth $TiO_2$-seeded MgO thin films as the tunnel barriers to alleviate the conductance mismatch problem[26,27]. The obtained nonlocal MR was 130 Ω (Fig. 1e), which corresponds to a spin injection efficiency of ~ 30%[6].

To date, there have been many reports on spin injection and transport in graphene in the nonlocal geometry[6,8,9,18-21,28-36]. Depending on the nature of the interface between graphene and the FM electrodes, we can sort them into three classes. The first class has pinhole contacts, in which the FM materials are directly contacting the graphene through tiny holes in an insulating barrier, such as evaporated $Al_2O_3$[24]. The second class has transparent contacts, in which the FM electrodes are directly contacted to graphene[23], or through a layer of nonmagnetic metal such as copper[28]. The third class has tunneling contacts, in which there is a thin insulting barrier between graphene and the FM electrodes. This includes $TiO_2$-seeded MgO barriers[6,9], unseeded MgO[8,36], hexagonal boron nitride (h-BN)[37], fluorinated graphene[35], and $Al_2O_3$ grown by atomic layer deposition[29]. With pinhole barriers, the nonlocal MR is around 10 Ω, corresponding to a spin injection efficiency of 2-18%. Although relatively high spin injection efficiency was also achieved in several pinhole contact devices, the reproducibility is poor, which might be due to the random nature of pinholes. With transparent contacts, the nonlocal MR is about several hundred mΩ with spin injection efficiency of ~1%. With tunnel barriers, the largest nonlocal MR is 130 Ω[6], and the highest spin injection efficiency is over 60%[35].

The interface between graphene and FM electrodes also plays an important role in the local MR measurements. After the first report of local MR on graphene in 2006[38], a detailed study of the interface between graphene and FM electrodes was performed and a strong correlation between good tunneling contacts and the observation of local MR signal was reported[39]. Generally speaking, it is much more difficult to observe spin transport in the local geometry due to the presence of charge current between the spin injector (E2) and the spin detector (E3), which produces a large spin-independent background signal. For example, in the 2007 study by van Wees' group, clean nonlocal spin signals are observed on several samples, but local measurements performed on the same samples exhibit local MR only on one of the samples and the signal is substantially noisier (Fig 1d). Recently, a large local MR of ~1 MΩ was observed on epitaxial graphene on SiC[10] (Fig. 1f).

Subsequently, spin injection and transport in large area graphene and high mobility graphene have been investigated. For instance, spin transport in large area graphene fabricated by chemical vapor deposition has been achieved, which is a key towards graphene spintronics at the wafer scale[31]. Furthermore, efforts have focused on spin transport in high mobility graphene by using suspended graphene[34,40] and graphene on h-BN[7]. These devices already exhibited longer spin diffusion lengths over several microns, although the quality was still below the devices used in charge transport. As the technology progresses, much better properties of graphene such as longer spin relaxation lengths and spin lifetime are observed[41,42]. In addition to the local and nonlocal measurements of spin transport in graphene, other techniques have been employed for spin injection and detection including spin pumping[43,44], three-terminal Hanle[45], and nonlinear spin detection with gold electrodes[46].

## 2. Magnetic Moments from Defects and Adatoms

The possibility of making graphene to be magnetic has attracted much interest, from both the basic scientific and technological standpoints. Scientifically, graphene does not consist of *d* or *f* electrons so the magnetic moment formation would be non-trivial. Technologically, making graphene magnetic could potentially give rise to high Curie temperature diluted magnetism and meet the demands of the ever increasing magnetic information storage density by engineering ultimately thin, 2D magnetic materials. Pristine graphene is strongly diamagnetic. The question of inducing magnetic ordering, or just magnetic moments as a first step, is of vital importance. The hope is to have tunable magnetism that could be changed by gating, doping, or functionalization. To date, there have been many theoretical and experimental studies of magnetic moment in graphene as a result of vacancy defects[47-4950,51], light adatoms[47,50,52-54] such as H and F, heavy adatoms[55,56] such as 3d, 4d and 5d elements, coupling to ferromagnetic substrates and molecular doping[57,58]. Magnetic moments are also predicted to form at the graphene edges[59], as shown in Fig. 2c. In addition, if the recent report of room-temperature ferromagnetism in hydrogenated epitaxial graphene on SiC[52] is confirmed, that would be an important step toward graphene applications in magnetic storage. In this article, we focus on the light adatom (H and F)-induced and defect-induced magnetic moments in graphene.

Theoretically, the existence of localized moments is often explained as a consequence of Lieb's theorem, derived for a half-filled single-band Hubbard model[60]. This theorem states that on a bipartite lattice the ground state has magnetic moment $\mu_B\,|N_A-N_B|$, where $N_A$ and $N_B$ are the numbers of sublattice sites. Effectively removing a site by placing on an adatom or creating a vacancy should then lead to a magnetic moment in the $\pi$ band, at least if the defect does not strongly couple $\pi$ and $\sigma$ bands.

Hydrogenated graphene is the benchmark case for graphene magnetism. Hydrogen chemisorbs reversibly on graphene, forming a strong covalent bond, effectively removing one $p_z$ orbital (shifts the bonding state down by several eVs) from the π band, thus creating a sublattice imbalance. The single hydrogen adatom induces a quasilocalized (resonant) state with magnetic moment of 1 $\mu_B$ (Bohr magneton) in accord with Lieb's theorem, see Fig. 2a. The existence of hydrogen-induced magnetic moment was first predicted theoretically[47]. First-principles calculations also show that doping of graphene could control the magnetic state in hydrogenated graphene[61]. In dense hydrogen coverage of single-side semi-hydrogenated graphene, the ground state appears to be an incommensurable spin spiral[62]. Another interesting light adatom is fluorine. It is in many cases similar to hydrogen: it bonds on top of a carbon atom[63], and transforms graphene to a wide gap insulator[64] with strong excitonic effects[65] at high fluorine coverages; fluorine can be reversibly chemisorbed on graphene[66]. The ground state of single-side semi-fluorinated graphene is predicted to be a 120° Néel antiferromagnetic state[62]. However, the question of whether an isolated fluorine induces a local magnetic moment is still open. On the one hand several experiments demonstrated that fluorination induces local magnetic moments[48,54,58] (see below). On the other hand, density functional theory results are inconclusive[67,68], mainly because of the self-interaction error in the exchange-correlation functionals that tends to delocalize electronic states[69,70].

A single vacancy in graphene generates a local spin-polarized electronic density by removing four electrons from the bands, see Fig. 2b. Three of those electrons form local $sp^2$ σ dangling bond states which split due to the crystal field and Jahn-Teller distortion. One state, deep in the Fermi sea, is doubly occupied. One, close to the Fermi level is occupied singly, contributing 1$\mu_B$ magnetic moment. The remaining electron is in the π quasilocalized state, forming a narrow resonant state[71]. By Lieb's theorem, the π state contributes 1 $\mu_B$ magnetic moment. Hund's coupling between the singly occupied σ and π states gives the total moment of 2 $\mu_B$. This value is predicted to be reduced to about 1.7$\mu_B$ due to the polarization of the itinerant π band, via resonant and Kondo-like coupling between the localized spin and itinerant band spins[51]. This scenario is consistent with density functional theory calculations[47,71], as well as with an experiment[48]. An important step to the goal of magnetism control has been demonstrating the switching off of the itinerant π part of the vacancy moments states by shifting the Fermi level, leaving the paramagnetic σ magnetic moments untouched. But a direct evidence of the π magnetism in the form of a spin-split peak in the scanning tunneling spectrum is still missing, although a profound resonance peak, which is a precursor for magnetic moment creation, has been observed[51]. Finally, whether or not DFT predicts a magnetic moment at an isolated vacancy in graphene is still a matter of debate[72].

Experimentally, to summarize, there are mainly four techniques for the detection of magnetic moments in graphene, as discussed in detail in the following.

The first one is using magnetization measurement via SQUID (superconducting quantum interference device). The basic mechanism of SQUID is the highly sensitive magnetic field dependence of the supercurrent in Josephson junctions. One example of this measurement is the detection of spin 1/2 paramagnetism in graphene with fluorine adatoms and vacancy defects[48]. In their experiments, the fluorination of the graphene was performed by exposing the graphene sample to $XeF_2$ at 200 $^oC$, while the vacancy defects were introduced by irradiation of graphene with protons and carbon ions. All the experimental results (the magnetic moments as a function of magnetic field) were fit by the Brillouin function:

$$M = NgJ\mu_B\left[\frac{2J}{2J+1}\operatorname{ctnh}\left(\frac{(2J+1)z}{2J}\right)-\frac{1}{2J}\operatorname{ctnh}\left(\frac{z}{2J}\right)\right] \quad (1)$$

where $z = gJ\mu_B H/k_B T$, $g$ is the g-factor, $J$ is the angular moment number, $N$ is the number of spins, and $k_B$ is the Boltzmann constant. By fitting the experimental results with different values of $J$, it was found that only $J$ = 1/2 provides a reasonable fit, while other $J$ could not. The magnetic moment was measured to be between 0.1-0.4 $\mu_B$ per vacancy (Fig. 2e) and 1 $\mu_B$per ~1000 fluorine adatoms, which suggests the cancellation of moments by fluorine clustering. However, no magnetic ordering was detected down to $T$=1.8 K[48].

The second method of detecting magnetic moments in graphene is via spin transport measurements. McCreary et al.[50] developed this method to detect the magnetic moment formation in hydrogen-doped graphene. The hydrogen doping was achieved by exposing graphene spin valve samples to atomic hydrogen at 15 K in an ultra high vacuum chamber and performing the spin transport measurements *in situ*. Interestingly, the nonlocal MR curves exhibit a dip centered at zero magnetic field after hydrogen exposure, which identifies magnetic moment formation in graphene (Fig. 2d). The basic mechanism is the spin scattering from exchange coupling with local magnetic moments. The sharpening of the Hanle curve after hydrogen exposure also indicates the formation of magnetic moments[73]. For vacancy defects introduced by argon sputtering, a similar behavior was observed[50].

The third method of detecting magnetic moments in graphene is via magnetic force microscopy or scanning tunneling microscopy[49,51]. One example is the identification of missing atoms as a source of carbon magnetism[51]. The single vacancies in these multilayer graphene were introduced by irradiation of low energy argon ions followed by high temperature annealing. From the differential conductance (*dI/dV*) spectra measured on a vacancy, a sharp resonance at the Fermi energy was observed. This resonance is a precursor for magnetic moment creation and is consistent with theoretical studies.

Another possible measurement of magnetic moments in graphene is based on the temperature dependence of the phase coherence length, such as the detection of local magnetic moments in fluorinated graphene[54]. In this study, the phase coherence length was measured as a function of temperature and carrier density. Both the phase coherence length and phase scattering time exhibited saturation at low temperatures (from 1 K to 10 K) at various carrier densities. Their results were consistent with spin flip scattering due to the local magnetic moments and it was claimed that magnetic moments could be formed by diluted fluorination.

## 3. Spin-Orbit Coupling

Spin-orbit coupling is an essential spin interaction[1,74]. On the one hand it is destructive of spin coherence, as it is usually responsible for spin relaxation. But spin-orbit coupling can also lead to many interesting phenomena, such as the spin Hall effect[75], topological quantum spin Hall effect[76], quantum anomalous Hall effect[77], spin-dependent Klein tunneling[78], weak antilocalization[79], or even strongly modifying the plasmon spectrum[80]. As with most graphene properties, one aims at tuning and controlling spin-orbit coupling by gating and functionalization.

Carbon is a light element with relatively weak spin-orbit coupling. The spin-orbit splitting of the 2p orbitals of the carbon ion is 7.86 meV[81]. In graphene, the spin-orbit coupling depends strongly on the bands, as well as on extrinsic effects such as adatoms, gating, and substrates. Without spin-orbit coupling the spectrum at the Fermi level forms Dirac cones. As soon as spin-orbit coupling is present, the cones separate, while preserving their spin degeneracy. In the presence of an external electric field transverse to graphene, the spin degeneracy is lifted by the Rashba effect[82], generating interesting band structure topologies, shown in Fig. 3. In Table I, we summarize the band structures at K and Γ for graphene structures as well as for selected low dimensional materials for comparison.

**Pristine graphene**

At the Γ point the spin-orbit splitting of the σ band is calculated to be 8.8 meV[83]. This coupling directly originates from the carbon spin-orbit coupling. At K, which is where the Fermi level lies, the spin-orbit splitting of the Dirac point is much less, in the range of 24-50 μeV[83-85], see Table I. This splitting comes from the hybridization of the $p_z$ orbitals, which form the Dirac cone, and d and higher carbon orbitals[83]. The energy spectrum of the Dirac electrons in graphene in the presence of spin-orbit coupling was first introduced by McClure and Yafet[86]. The McClure-Yafet Hamiltonian in modern notation reads

$$H = \lambda_I \kappa \sigma_z s_z \tag{2}$$

Here $\lambda_I$ is the intrinsic spin-orbit coupling parameter, κ is 1 for K and -1 for K', $\sigma_z$ is the pseudospin (sublattice space A and B) Pauli matrix, and $s_z$ is the (real) spin Pauli matrix. This intrinsic spin-orbit coupling opens a gap in the Dirac spectrum of the magnitude $\Delta_{SOC} = 2\lambda_I$. In the presence of the intrinsic spin-orbit coupling the electron states remain doubly degenerate (Fig. 5), due to the combination of time reversal and space inversion symmetry[74]. Based on the above Hamiltonian, Kane and Mele[76] predicted the quantum spin Hall effect in graphene, a precursor system for topological insulators. Unfortunately, the rather small value of the coupling makes its direct experimental observation a challenge. Rippled graphene should have larger spin-orbit coupling, due to the hybridization of the $p_z$ orbitals with $p_x$ and $p_y$ orbitals from the σ band[87]. Although relatively large values of spin-orbit coupling can be found in nanotubes[87,88] (see Table I), in rippled graphene the effect is weaker due to the alternating curvature. In the extreme case of a miniripple (A atoms shifted up, B down with respect to the graphene plane), the spin-orbit gap is still only 100 μeV for an 8% shift with respect to the lattice constant[83]. This is roughly the value estimated for the renormalization of the spin-orbit gap due to zero-point flexural motion[89,90]. Such values are at the current experimental capabilities of resolving the upper limit on the possible Dirac point bandgap[91].

Of other two dimensional materials, silicene and germanene are more promising in displaying spin-orbit effects, due to much larger values of spin-orbit couplings[92]. Insulating BN has a rather weak spin-orbit coupling[93], while the semiconducting family of $MoS_2$ materials have the conduction electron spin-orbit splitting at K reaching tens of meV[94-96]. See Table I for further details.

**Rashba effect in graphene**

If space inversion symmetry of graphene is broken by the substrate, external electric field, or adatoms, the Rashba effect appears[82], manifested by the spin splitting of doubly degenerate pristine graphene energy bands (Fig. 3). The Rashba effect due to an external field is calculated to be rather weak, 5 μeV for a field of 1 V/nm[83]. The corresponding Hamiltonian reads

$$H_R = \lambda_R(\kappa\sigma_x s_y - \sigma_y s_x) \qquad (3)$$

where $\lambda_R$ is the Rashba coupling. Different from conventional semiconducting two dimensional electron gases in which the Rashba coupling is linear in the momentum, the Rashba coupling in graphene does not depend on the momentum. The reason is that Rashba coupling is proportional to velocity which is constant for massless Dirac electrons in graphene. Microscopically, the Rashba coupling in graphene comes from the $\pi - \sigma$ hybridization[97]; $d$ orbitals play almost no role here[98]. If the spin-inversion symmetry is broken locally, say, due to adatoms, the Rashba coupling (but also the intrinsic one, see below) become space dependent: $\lambda_R = \lambda_R(x, y)$.

**Bilayer, trilayer, graphite**

Interlayer coupling does not affect in any essential way the spin-orbit interaction (see Table I), which comes mainly from the core region of the atoms. First-principles investigations of bilayer[99] and trilayer graphene[100], as well as graphite[100] show that the spin-orbit coupling effects come almost solely from the intralayer spin-orbit coupling $\lambda_I$. This conclusion is also supported by multiband tight-binding investigations of trilayer graphene[101].

**Light adatoms**

In order to observe spin-orbit effects in graphene, it appears necessary to enhance the spin-orbit coupling, without destroying much of the Dirac cone structure. A viable route is adding light adatoms which, even in a rather dilute limit, could significantly enrich the spin-orbit physics of graphene. Adatoms, such as hydrogen, can induce giant spin-orbit coupling of the Dirac electrons by locally producing $sp^3$ bonding. The bonding brings the $\sigma$ states to the $\pi$ band. Since the $\sigma$ states have spin-orbit coupling of about 10 meV, as in the carbon atom, this is the limiting value we can expect for such an enhancement[102]. Heavier adatoms can produce even stronger spin-orbit coupling[56,103], coming from their own orbitals, but at the expense of more radically changing the local electronic structure.

Similarly to the magnetic moment case, also for spin-orbit coupling, hydrogen is a benchmark adatom. Its s orbitals form a covalent bond with carbon p orbitals, locally distorting the lattice towards tetrahedral $sp^3$ bonding. This leads to a locally induced spin-orbit coupling. In addition to the intrinsic and Rashba, a new spin-orbit hopping term due to pseudospin inversion asymmetry (PIA) emerges[104]. First-principles calculations have shown that the enhancement of the spin-orbit coupling (of all three types) is indeed giant, of the order of 1 meV, about two orders more than in pristine graphene[104]. A recent experiment on the spin Hall effect in hydrogenated graphene has observed such an enhancement[75]. In Table I the spin-orbit splittings are shown for the single side semi-hydrogenated graphene.

<table>
<tr><th>Material</th><th>Structure</th><th>Spectrum at K</th><th>$\Delta_K$</th><th>Spectrum at Γ</th><th>$\Delta_\Gamma$</th></tr>
<tr><td>Graphene</td><td>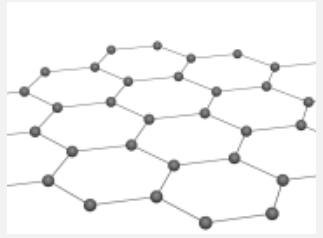</td><td>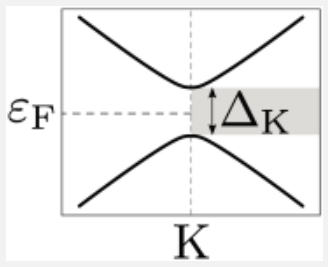</td><td>24 μeV[83,84] - 50 μeV[85]</td><td>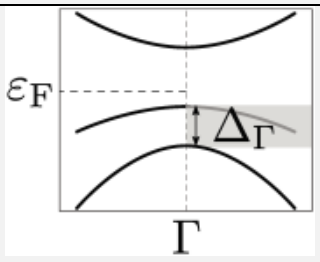</td><td>8.8 meV</td></tr>
<tr><td>bilayer graphene</td><td>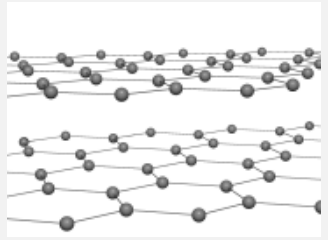</td><td>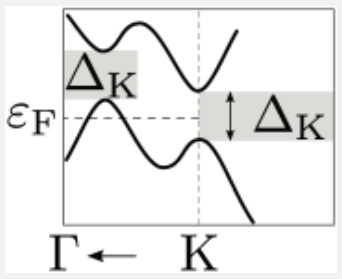</td><td>24 μeV[99]</td><td>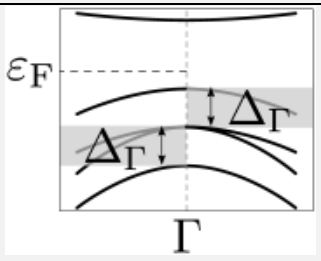</td><td>8.6 meV</td></tr>
<tr><td>armchair nanotube (4,4)</td><td>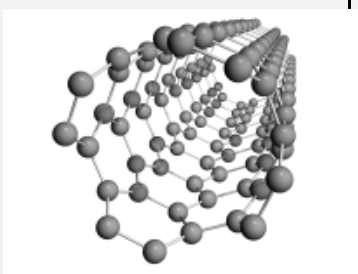</td><td></td><td></td><td>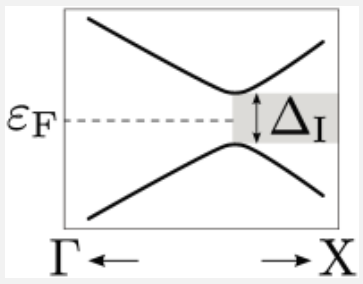</td><td>1.6 meV</td></tr>
<tr><td>Single-side semi-hydrogenated graphene</td><td>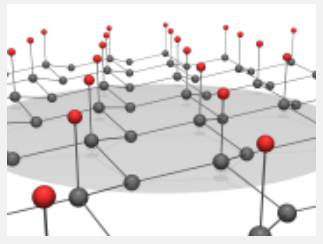</td><td>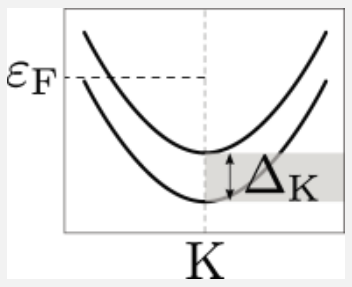</td><td>0.2 meV[104]</td><td>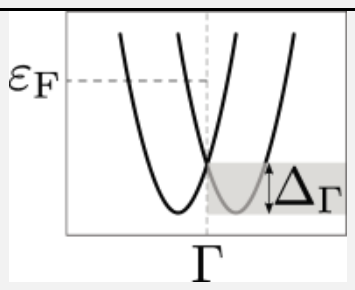</td><td>0.8 μeV[104]</td></tr>
<tr><td>Silicene</td><td>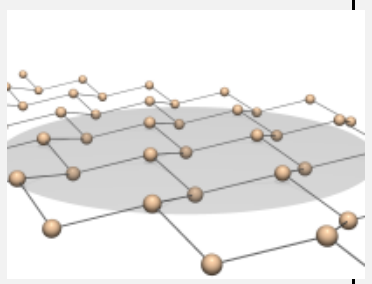</td><td>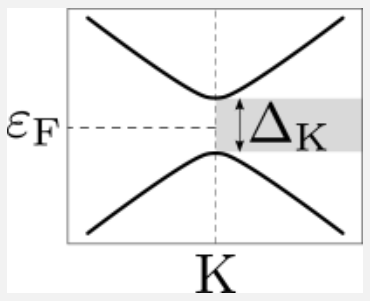</td><td>1.6 meV[92]</td><td>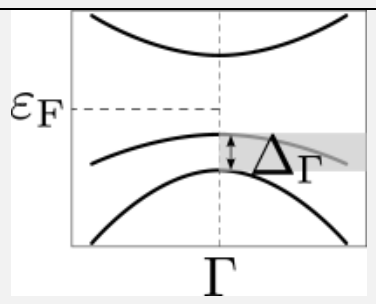</td><td>33.7 meV</td></tr>
<tr><td>Germanene</td><td>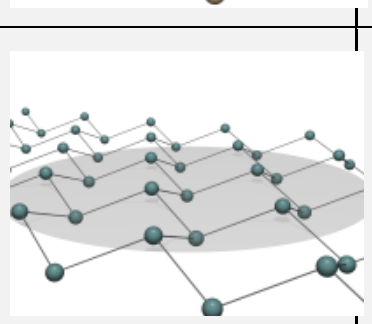</td><td>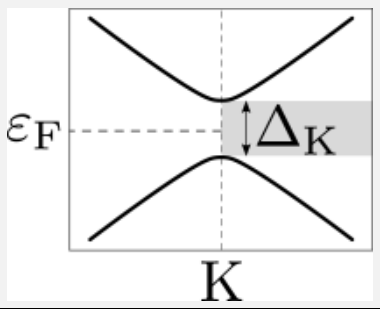</td><td>24 meV[92]</td><td>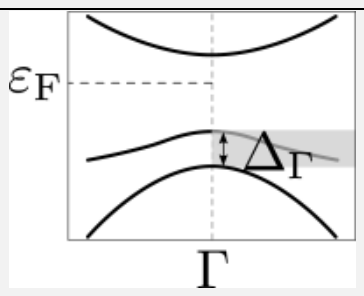</td><td>195 meV</td></tr>
<tr><td>2D hexagonal BN<br>$\varepsilon_g = 6$ eV[93]</td><td>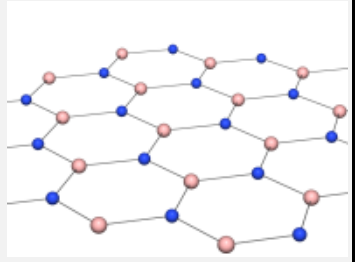</td><td>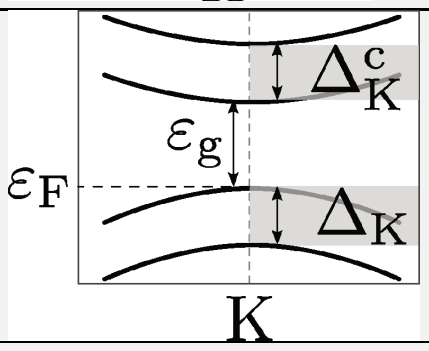</td><td>30 μeV<br>15 μeV</td><td>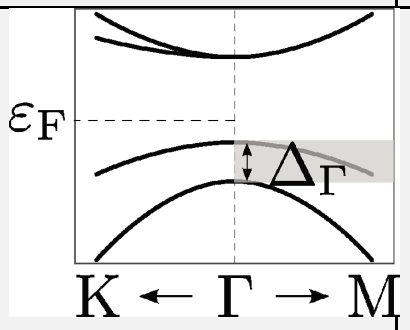</td><td>12.6 meV</td></tr>
<tr><td>2D $MoS_2$<br>$\varepsilon_g = 2.89$ eV[94]</td><td rowspan="4">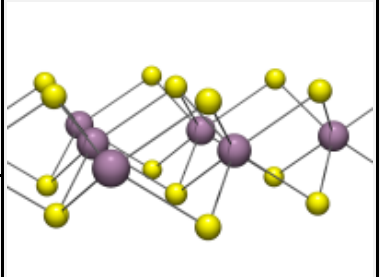</td><td rowspan="4">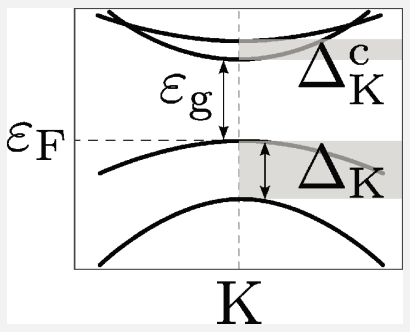</td><td>3 meV[95]<br>147 meV[95]</td><td rowspan="4">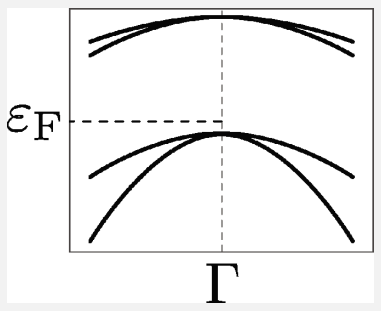</td><td>0</td></tr>
<tr><td>2D $MoSe_2$[96]<br>$\varepsilon_g = 2.4$ eV</td><td>21 meV[95]<br>186 meV[95]</td><td>0</td></tr>
<tr><td>2D $WS_2$[96]<br>$\varepsilon_g = 3.1$ eV</td><td>27 meV[95]<br>433 meV[95]</td><td>0</td></tr>
<tr><td>2D $WSe_2$[96]<br>$\varepsilon_g = 2.7$ eV</td><td>38 meV[95]<br>463 meV[95]</td><td>0</td></tr>
</table>

**Table I. Spin-orbit coupling in low-dimensional materials**

**Table I.** Spin-orbit gaps $\Delta_K$ at K and $\Delta_\Gamma$ at Γ points of the Brillouin zone for selected low-dimensional systems. Also sketched are the corresponding band structures. $\varepsilon_F$ is the Fermi energy and $\varepsilon_g$ is the bad gp. The color codes used in the structure diagrams are: grey, C; red, H; orange, Si; dark green, Ge; blue, B; pink, N; purple, Mo/W; yellow, S/Se. If two values are shown, the upper is for the conduction, the lower for the valence band. Unless referenced, the values were calculated from first principles[152] for the purposes of this review.

## 4. Spin Relaxation in Graphene

The major scientific issue within graphene spintronics is the large discrepancy between the theoretical and experimental values of spin lifetime in graphene. As we will show, theory predicts spin lifetimes of ~1 μs for pristine graphene, while experiment measures values ranging from tens of ps to a few ns[5-9,24,34,36,40,45,105-107]. In the best case, this is a discrepancy of over two orders of magnitude. This suggests that the source of spin relaxation is of extrinsic origin (e.g. impurities, defects, static ripples, etc.) and the challenge is to identify the microscopic mechanism.

Experimental studies of spin relaxation are typically based on Hanle spin precession measurements on graphene spin valves in the nonlocal geometry. This is performed by applying an out-of-plane magnetic field (perpendicular to the graphene, as shown in figure 4a), which causes the spins in graphene to precess as they diffuse from the spin injector (E2) to the spin detector (E3). Figure 4b shows typical Hanle spin precession curves, which were obtained by measuring the nonlocal resistance as a function of the out-of-plane magnetic field ($B_\perp$) for a graphene spin valve with tunneling contacts. The top branch (red curve) is for the parallel magnetization state of E2 and E3, and the bottom branch (black curves) is for the antiparallel magnetization state. The characteristic reduction in the nonlocal resistance with increasing magnitude of $B_\perp$ is a result of spin-precession, which reduces the spin polarization reaching the detector electrode. For the case of highly resistive contacts (i.e. contact resistance, $R_C$, is much larger than the spin resistance of graphene, $\rho\lambda_s/W$, where $\rho$, $\lambda_s$, and $W$ are the resistivity, spin diffusion length, and the width of graphene, respectively), the nonlocal Hanle curve is given by the relation[1,14,108]:

$$R_{NL} \propto \pm \int_0^\infty \frac{1}{\sqrt{4\pi Dt}} \exp\left[-\frac{L^2}{4Dt}\right] \cos(\omega_L t) \exp(-t/\tau_s) dt \quad (4)$$

where the + (-) sign is for the parallel (antiparallel) magnetization state, $L$ is the distance from injector (E2) to detector (E3), $D$ is the diffusion constant, $\tau_s$ is the spin lifetime, and $\omega_L = g\mu_B B_\perp/\hbar$, in which $g$ is the g-factor, $\mu_B$ is the Bohr magneton, and $\hbar$ is the reduced Planck's constant. Intuitively, the factors containing $D$ represent the diffusion of the spin-polarized electrons, the factor containing $\omega_L$ represents the spin precession, the factor containing $\tau_s$ represents the spin relaxation, and the integration is performed over the distribution of transit times for spins to diffuse from injector (E2) to detector (E3). Using this equation, we fit the parallel and anti-parallel Hanle curves (solid lines) assuming $g$ = 2. The fitting parameters obtained are $D$ =2.0×10$^{-2}$ m$^2$s$^{-1}$ and $\tau_s$ = 771 ps, which corresponds to $\lambda_s = \sqrt{D\tau_s}$ = 3.9 μm.

In practice, one must be careful when analyzing Hanle curves. In the case of lower contact resistances, equation 4 is not valid and one must use a more complicated analysis that takes into account the contact-induced spin relaxation due to escape time effects[24,109]. While this analysis has traditionally required the numerical solution to a set of linear equations, an analytic closed-form solution has recently been obtained[110]. One should also be cautious when the g-factor could deviate from 2. For instance, the presence of paramagnetic moments generates an effective exchange field and results in faster spin precession and enhanced effective g-factor. In this case, it becomes difficult to distinguish between faster spin precession due to magnetic moments and longer spin lifetime in graphene, so an alternative analysis of the Hanle curve is required[50,73,111].

Experimentally, spin lifetimes up to a few ns have been observed for graphene spin valves on $SiO_2$ with tunneling contacts[6,8,9]. On the other hand, graphene spin valves on $SiO_2$ with pinhole contacts of similar contact resistance exhibit lower spin lifetimes (typically around 200 ps). Therefore, in addition to escape time effects[24,109], other types of contact-induced spin relaxation effects could play an important role in the Hanle measurement. For example, ferromagnetic contacts could introduce spin relaxation through a number of mechanisms, including through the inhomogeneous magnetic fringe fields, and interfacial spin scattering between the FM and graphene. Thus, to investigate spin relaxation in graphene, it is crucial to minimize contact-induced spin relaxation effects by making the distance between FM electrodes much larger than the bulk graphene spin relaxation length, and improving the quality of the contacts.

Theoretically, there have been numerous studies[112-115], but the origin of spin relaxation in graphene has remained elusive. Two mechanisms of spin relaxation have been widely applied in an effort to explain experimental trends in graphene: Elliott-Yafet[116,117] (EY) and Dyakonov-Perel[118] (DP) mechanisms. Both have their roots in metal and semiconductor spintronics[74,119]. Both mechanisms rely on spin-orbit coupling and momentum scattering, but their effect is opposite with respect to the latter. The EY mechanism explains the spin relaxation by spin flips during scattering (Fig. 5). In the presence of spin-orbit coupling the Bloch states are an admixture of the Pauli spin up and spin down states. Say, a spin 'up' electron has a small admixture amplitude of the Pauli spin down spinor. Then even a nominally spin conserving impurity or phonon scattering can induce a spin flip. Typically an electron undergoes up to a million scattering events before its spin is flipped. The spin relaxation rate is roughly

$$1/\tau_s \approx b^2/\tau_p \tag{5}$$

where $\tau_p$ is the momentum relaxation time and $b$ is the amplitude of the spin admixture. Typically $b$ is determined as the intrinsic spin orbit coupling divided by the Fermi energy ($\varepsilon_F$), $b \approx \lambda_I/\varepsilon_F$ . If we take representative values for graphene flakes used in experiments, $\lambda_I \sim 10$ μeV, $\varepsilon_F \sim 100$ meV, and $\tau_p \sim 10$ fs, the spin relaxation time comes to $\tau_s \approx 1$ μs. This is more than two orders magnitude than even the longest values observed experimentally.

The DP mechanism is based on the concept of motional narrowing: the more the electron scatters, the less its spin relaxes (the more narrow would be the resonance line in a spin resonance experiment); see Fig. 5. Unlike for the EY mechanism, the spins precess between the scattering events. In the absence of space inversion symmetry, the spin-orbit coupling is manifested as a spin-orbit field, say of the Rashba type. The electron spin precesses along this field. As the electron scatters, the orientation (and/or the value) of the effective magnetic field changes. In effect, the electron spin precesses in a randomly fluctuating spin-orbit field, with the correlation time of the fluctuations given by the momentum relaxation time. The spin relaxation rate is then given by

$$1/\tau_s \approx \lambda_R^2 \tau_p \tag{6}$$

where $\lambda_R$ is the magnitude of the Rashba field, which is the spin precession frequency. Typically $\lambda_R \sim 1$ μeV and $\tau_p \sim 10$ fs, giving the relaxation time $\tau_s \approx 1$ μs, as large as that coming from the EY mechanism. Recently, another mechanism based on the spin-pseudospin entanglement has been proposed as a highly efficient spin relaxation process in the ballistic limit[120].

Pristine graphene should have its spin relaxation limited by the intrinsic spin-orbit coupling and phonons. Particularly efficient are flexural modes which provide large enhancement of spin-orbit coupling due to $\pi - \sigma$ mixing. This essentially Elliott-Yafet mechanism can lead to spin relaxation times at room temperature in the range of microseconds to tens of nanoseconds, depending on how the amplitude of the flexural modes is limited, as well as a marked temperature dependence and anisotropy[90,115]. This pristine limit has not yet been experimentally realized.

Experimental studies have been performed to investigate the relative importance of the EY and DP mechanisms. For single layer graphene, when the carrier density is tuned by backgate, it is observed that the $\tau_s$ increases with the diffusion constant $D$ ($\sim \tau_p$), which is consistent with the EY spin relaxation mechanism (Fig. 4c) [9,24,105]. For bilayer graphene, it was found that $\tau_s$ decreases with increasing $D$ (or $\tau_p$), which suggests a DP spin relaxation mechanism (Fig. 4d) [8,9]. On the other hand, there are also reports suggesting EY spin relaxation mechanisms in few layer graphene[121] and DP spin relaxation mechanism in single layer graphene[36]. The roles of the EY and DP spin relaxation mechanisms remain an open question[122,123]. For example, it could be hard to distinguish these mechanisms in graphene because both EY-like and DP-like behaviors could be observed due to the randomness of the Rashba field. Recent experimental studies of spin relaxation in high quality graphene, such as graphene on h-BN[7], epitaxial graphene[111] or suspended graphene[34,40] (Fig. 4e), exhibit similar spin lifetimes compared to exfoliated graphene on $SiO_2$. These indicate that mobility is not the limiting factor for spin lifetime. Along these lines, Han et al studied the spin lifetimes in single layer graphene spin valves with tunable mobility, using organic ligand-bound nanoparticles as charge reservoirs[107]. At fixed carrier concentration (Fig. 4f), it was observed that spin lifetime exhibits little variation as mobility is tuned between 2,700 and 12,000 $cm^2/Vs$. These results demonstrate that charged-impurity scattering is not primarily responsible for generating spin relaxation in graphene. Furthermore, a recent experimental study demonstrated the negligible effect of hyperfine interaction even in isotopically-engineered $^{13}C$ graphene, which has a high density of nuclear moments[124]. But the question remains, what is the primary source of spin relaxation?

Recent experiments of universal conductance fluctuations and weak localization of nominally pristine graphene on $SiO_2$ have found that a substantial contribution to the dephasing rate is due to spin flip scattering by magnetic moments[125]. This has motivated another proposal for the mechanism of spin relaxation in graphene[126] which seems to explain quantitatively the available experimental data. It is based on the existence of magnetic moments (from vacancies or adatoms) that act as resonant scatterers. Magnetic scatterers provide the spin-flip exchange field. At resonant energies, the scattering electron spends a considerable time at the impurity, allowing the electron spin to precess, say, at least a full circle (Fig. 5). As the electron escapes, the spin can be found with equal probability up and down, so the spin flip time equals the spin relaxation time. This happens precisely at the resonant energy. Averaging over resonant energies (due to thermal fluctuations, distributions of resonant energies of different defects, or due to the presence of electron-hole puddles), can yield the spin relaxation times of 100 ps[126] for as little as 1 ppm of magnetic moments.

Determining the origin of spin relaxation in graphene remains an important challenge. Experiments on the injector-detector spacing dependence of spin lifetimes could help elucidate the role of contacts. Investigating the anisotropy of spin relaxation (the in-plane spin lifetime of in-plane, $\tau_{s\parallel}$, and the out-of-plane spin lifetime, $\tau_{s\perp}$) could contribute to the identification of relevant spin-orbit mechanisms[42] ($\tau_{s\parallel} = 2\tau_{s\perp}$ for DP; $\tau_{s\parallel}/\tau_{s\perp} \sim 0$ for EY from phonon scattering;

no particular relation for EY from impurity scattering). The effects of hydrogen adatoms also needs further investigation because their effect appears to highly depend on the synthesis methods: low temperature atomic hydrogen produces magnetic moments[50], hydrogen plasma results in enhanced spin lifetimes[127], while HSQ irradiation generates large spin-orbit coupling[75]. Identifying the microscopic spin relaxation mechanism will make it possible to determine a course of action for increasing the spin lifetime toward the theoretical limit, which will have important implications for basic science and technological applications.

## 5. Potential Applications and Future Directions

Recent experimental studies have already identified the advantages of graphene for spintronics compared to metals and semiconductors. Table II lists the spin dependent properties (spin lifetime, spin diffusion length, and spin signal) of graphene, Cu, Ag, Al, and doped semiconductors, including Si, GaAs and Ge, obtained by nonlocal and local spin transport measurements[6,8-10,15,108,128-132]. The room temperature physical characteristics including long spin diffusion length, large spin signal, and relatively long spin lifetime (which could be extended further) make graphene one of the most favorable candidates for spin channel material in spin logic applications.

| Spin Channel | | Spin lifetime | Spin diffusion lengths | Spin signals |
|---|---|---|---|---|
| Metals | Cu[15,131] | ~ 42 ps at 4.2 K<br>~ 11 ps at 300 K | ~ 1 μm at 4.2 K<br>~ 0.4 μm at 300 K | ~ 1 mΩ at 4.2 K<br>~ 0.5 mΩ at 300 K |
| | Al[108] | ~ 100 ps at 4.2 K<br>~ 45 ps at 300 K | ~ 0.6 μm at 4.2 K<br>~ 0.4 μm at 300 K | ~ 12 mΩ at 4.2 K<br>~ 0.5 mΩ at 300 K |
| | Ag[132] | ~ 20 ps at 5 K<br>~ 10 ps at 300 K | ~ 1 μm at 5 K<br>~ 0.3 μm at 300 K | ~ 9 mΩ at 5 K<br>~ 2 mΩ at 300 K |
| Semiconductor | Highly doped Si[129,153] | ~10 ns at 8 K<br>~1.3 ns at 300 K | ~2 μm at 8 K<br>~0.5 μm at 300 K | ~ 30 mΩ at 8 K<br>~ 1 mΩ at 300 K |
| | GaAs[154] | 24 ns at 10 K<br>4 ns at 70 K | 6 μm at 50 K | ~ 30 mΩ at 50 K |
| | Highly doped Ge[130] | ~ 1 ns at 4 K<br>~ 300 ps at 100 K | ~ 0.6 μm at 4 K | 0.1-1 Ω at 4 K<br>0.02 ~ 0.1 Ω at 200 K |
| Graphene[6,9,10] | | 0.5 - 2 ns at 300 K<br>1 - 6 ns at 4 K | 3 - 10 μm at 300 K<br>(~100 μm fit from local MR data) | 130 Ω at 300 K<br>(1 MΩ for local MR at 1.4 K) |

**Table II**. Spin dependent properties of graphene, metals and semiconductor measured by spin valve measurements. To be noted, other methods have been employed to measure spin lifetimes in semiconductors. Spin lifetimes in intrinsic Si and Ge were obtained using the ballistic hot electron injection method. The spin lifetime of intrinsic Si is about 10 ns at room temperature, and ~ 1 microsecond at low temperature[155], and the spin lifetime of intrinsic Ge, is between 100 ns to 400 ns in the temperature range 30K -60 K[156]. Optical pump-probe methods have measured spin lifetimes of ~100 ns at 5 K in lightly doped n-GaAs[157].

Graphene-based spintronic devices for logic applications have been proposed, such as graphene spin logic by Dery and Sham[2,3], and all spin logic in Behin-Aein et al[133]. The building block of graphene spin logic is a magnetologic gate (MLG) consisting of a graphene sheet contacted by five ferromagnetic electrodes (Fig. 6a). Two electrodes (A and D) define the input states, two electrodes (B and C) define the operation of the gate, and one electrode (M) is utilized for readout. The logic operation is performed by spin injection/extraction in graphene at the input/operation electrodes (A,B,C,D) followed by mixture and diffusion of spin currents to the output (M). The magnetization state of the electrodes could be controlled based on spin-transfer torque in the metallic FM electrodes. Fig. 6b shows the schematic of all spin logic. The information is stored in the bistable states of magnets. Corresponding inputs and outputs communicate with each other via spin currents through a spin transport channel, such as graphene. The final state is written to the output magnet by spin-transfer-torque switching from the pure spin diffusion current. While this has not been demonstrated experimentally, a magnetic-field-assisted spin torque switching has been performed recently[134]. Both approaches of spin logic are based on pure spin diffusion currents (i.e. nonlocal) and have demanding technical challenges including the development of more robust and uniform tunnel barriers for spin injection, the integration of half-metallic ferromagnets and perpendicular ferromagnets, increased spin diffusion lengths, current-based spin detection[135], and spin-amplification in confined geometries (device size $<< \lambda_S$)[136].

Beyond these existing proposals, there is an opportunity to develop novel physical properties by introducing exchange fields and spin-orbit coupling in graphene. For example, novel topological phases in graphene were proposed, including the topological quantum spin Hall effect and the quantum anomalous Hall effect[56,76,77,103,137,138]. Gate-tunable exchange fields from adjacent ferromagnetic insulators are proposed as spin-rotators to manipulate the spin polarization within graphene[139-141]. Similarly, enhanced spin-orbit coupling by adatom doping and/or overlayer growth could be used to manipulate spins in graphene. As discussed earlier, hydrogen adatoms have produced magnetic moments and resulting exchange fields[50], as well as enhanced spin-orbit coupling[104] and resulting spin Hall effect[75]. Recent success in the growth of ferromagnetic insulator EuO[73,142], topological insulator candidate $Bi_2Se_3$[143] and transition metal dichalcogenide $MoS_2$[144] on graphene opens up this promising direction. Beyond graphene, emerging 2D materials[145] offer different properties that may be useful for spintronic devices. Hydrogen-terminated germanane[146,147] and monolayer transition metal dichalcogenides such as $MoS_2$[148,149] are semiconductors with direct band gap and higher spin-orbit coupling compared to graphene (see Table I). This enables optical coupling to the spin degree of freedom, as well as strong spin Hall effects[150]. Meanwhile, stannanane (i.e. 2D Tin) and $sp^2$ germanene are predicted to generate topological quantum spin Hall effect at elevated temperatures[92,151]. Furthermore, stacked heterostructures integrating these emerging materials with graphene could open up new possibilities, taking advantage of the long spin diffusion lengths in graphene and the novel physical properties in the emerging materials for spin manipulation. These new behaviors could be exploited to develop new device concepts.

In conclusion, tremendous progress has been made in graphene spintronics over the past several years. Looking forward, there are many more opportunities. The origin of spin relaxation in graphene is still a major open question, and progress toward long spin lifetimes and spin diffusion lengths are important for graphene-based spintronic devices. In addition, the nature of magnetic interactions in graphene has scarcely been explored. Furthermore, the development of graphene hybrid structures and alternative 2D materials should generate new spin dependent physical properties and novel devices due to enhanced spin-orbit and exchange interactions.

## Acknowledgement

W. H. and R. K. acknowledge the support of ONR (N00014-14-1-0350), NSF (DMR-1310661), and NSF-MRSEC (DMR-0820414), NRI-NSF (NEB-1124601) and ARO (W911NF-11-1-0182). R. K. also acknowledges the support from C-SPIN, one of six centers of STARnet, a Semiconductor Research Corporation program sponsored by MARCO and DARPA. J. F. and M.

G. acknowledge support by the DFG SFB 689, SPP 1285, and GRK 1570. J. F. also acknowledges support from EC under Graphene Flagship (Contract No. CNECT-ICT-604391).

## Figure captions

**Figure 1. Spin injection and transport in graphene spin valves. a-b**, nonlocal (a) and local (b) spin transport measurement geometries. Blue symbols indicate spin polarized carriers. **c-d**, typical nonlocal (c) and local (d) MR curves measured on graphene with $Al_2O_3$ barriers. The arrows indicate the magnetization directions of four ferromagnetic electrodes. Inset is the schematic of the device geometry. **e**, large nonlocal MR measured on graphene spin valves with tunneling contacts at $V_g = 0$ V. Black/red traces indicates the data measured during the sweeping up/down the magnetic field. Inset is the schematic of the device measurement geometry. **f**, large local MR measured on epitaxial graphene grown on SiC with highly resistive tunneling contacts. Fig. c and Fig. d, reproduced from ref. 5. Fig. e, reproduced from ref. 6. Fig. f, reproduced from ref. 10.

**Figure 2. Magnetic moment in graphene due to light adatoms and vacancy defects. a-c,** theoretical prediction of magnetic moment in graphene due to hydrogen (a), vacancy defects (b), and at the graphene edges (c). Red and blue denote the opposite spin polarizations. **d**, the magnetic moments due to hydrogen doping detected by spin transport measurements at 15 K. The device was measured after 8 s hydrogen doping. Black line is the experimental result, and the red curve is fitting based on the spin scattering model due to local magnetic moments. The inset shows a schematic image of spin (black arrow) scattered by local magnetic moments (green arrow). **e,** the magnetic moments due to vacancy defects detected by SQUID. Error bars indicate the accuracy of determination of the number of spins per vacancy. Inset: Magnetic moment due to vacancies as a function of parallel field H. The solid lines are fitted curve based on Brillouin function with $J = 1/2$. Fig. d reproduced from ref. 50 and Fig. e reproduced from ref. 48.

**Figure 3. Band structure topologies of graphene with spin-orbit coupling in a transverse electric field.** Touching Dirac cones exist only when spin-orbit coupling is neglected (first from left). As long as it is present, the orbital degeneracy at the Dirac point is lifted and the spin-orbit gap appears (second from left). In an external electric field perpendicular to graphene, due to a gate or a substrate, the Rashba effect lifts the remaining spin degeneracy of the bands (third, fourth and fifth from left). If the intrinsic and Rashba couplings are equal, at a certain value of the electric field, two bands (red) form touching Dirac cones again (fourth from left). If the Rashba coupling dominates (fifth from left), the spin-orbit gap closes. Red and blue denote the opposite spin polarizations.

**Figure 4. Experimental studies of spin relaxation in graphene. a,** schematic of Hanle measurement by applying an out of plane magnetic field ($\mathbf{B}_{\perp}$, blue arrow). Black arrows indicate the magnetization direction of the ferromagnetic electrodes. Red arrows indicate the spin orientation processing due to the existence of the magnetic field. **b,** typical Hanle curves in graphene spin valves with tunneling contacts. The red (black) circles are data for parallel (antiparallel) alignment of the central electrodes. The red and black lines are curve fits based equation 4. The fitted spin lifetime is 771 ps and diffusion constant is 0.020 $m^2/s$. **c-d**, spin lifetime (squares) and diffusion coefficient (circles) as a function of gate voltage at 4 K for single layer graphene (c) and bilayer graphene (d) respectively. Error bars represent the 99% confidence interval. **e**, Hanle curves measured on suspended graphene spin valves. Black curves are the experimental results and the red line is the fitting. Inset is the scanning electron micrographs of a typical suspended graphene device. **f**, spin lifetime as a function of carrier densities for same graphene spin valve measured at 10 K with tunable mobility: 4200 $cm^2$/Vs (Black squares), 12000 $cm^2$/Vs (red circles) and 4050 $cm^2$/Vs (black open squares). Inset, the graphene spin valve device geometry with organic ligand-bound nanoparticles. Error bars represent the 99% confidence interval. Fig. c and Fig. d reproduced from ref. 9. Fig. e reproduced from ref. 40. Fig. f, reproduced from ref. 106.

**Figure 5. Spin relaxation mechanisms in graphene.** An illustrative figure of three possible spin relaxation mechanisms for graphene: Elliott-Yafet, Dyakonov-Perel, and resonant scattering by local magnetic moments. The blue dots indicate the electrons/holes with yellow arrows as their spin orientation. The red dots represent the scattering centers. The grey cones with circular arrows represent the spin precession.

**Figure 6. Spin logic application of graphene spin valves. a,** schematic drawing of graphene based magnetologic gate (MLG) consisting of a graphene sheet contacted by five ferromagnetic electrodes. Two electrodes (A and D) define the input states, two electrodes (B and C) define the operation of the gate, and one electrode (M) is utilized for readout. $V_{dd}$ drives the steady-states current, and $I_M(t)$ is the digital transient current response, which gives the output. $C_M$ is a capacitor. $I_W$ is the writing current to manipulate the magnetization direction of each ferromagnetic electrode. $I_r(t)$ is the current used to perturb the magnetization of the electrode M. **b**, "All spin logic" proposed by Behin-Aein et al. 1, 2, 3, 4 and 5 are magnetic free layer, isolation layer, tunneling layer, channel/interconnect, and the contact respectively. Fig. b, reproduced from ref. 133.

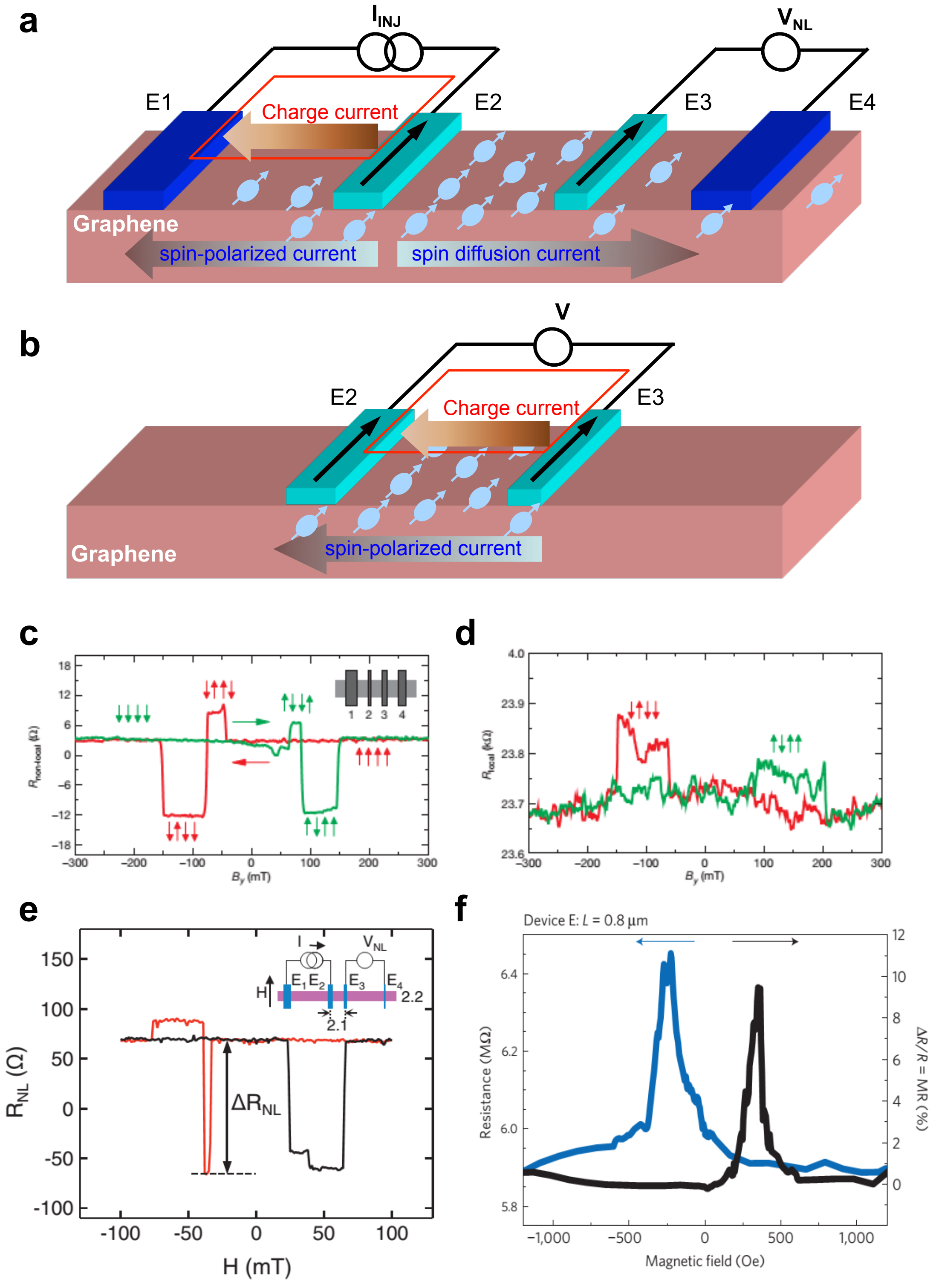

**Figure 1. Spin injection and transport in graphene spin valves.**

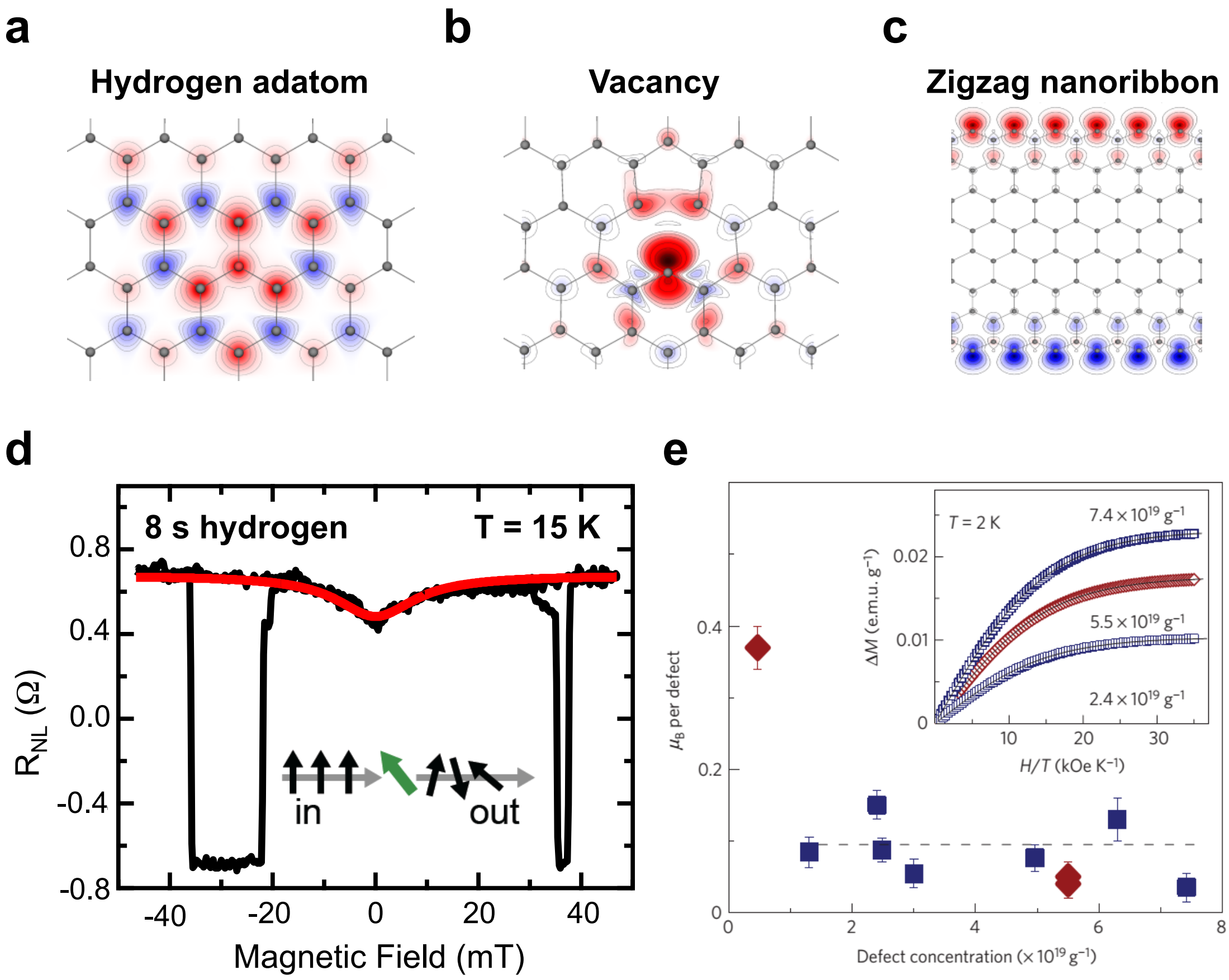

**Figure 2. Magnetic moment in graphene due to light adatoms and vacancy defects.**

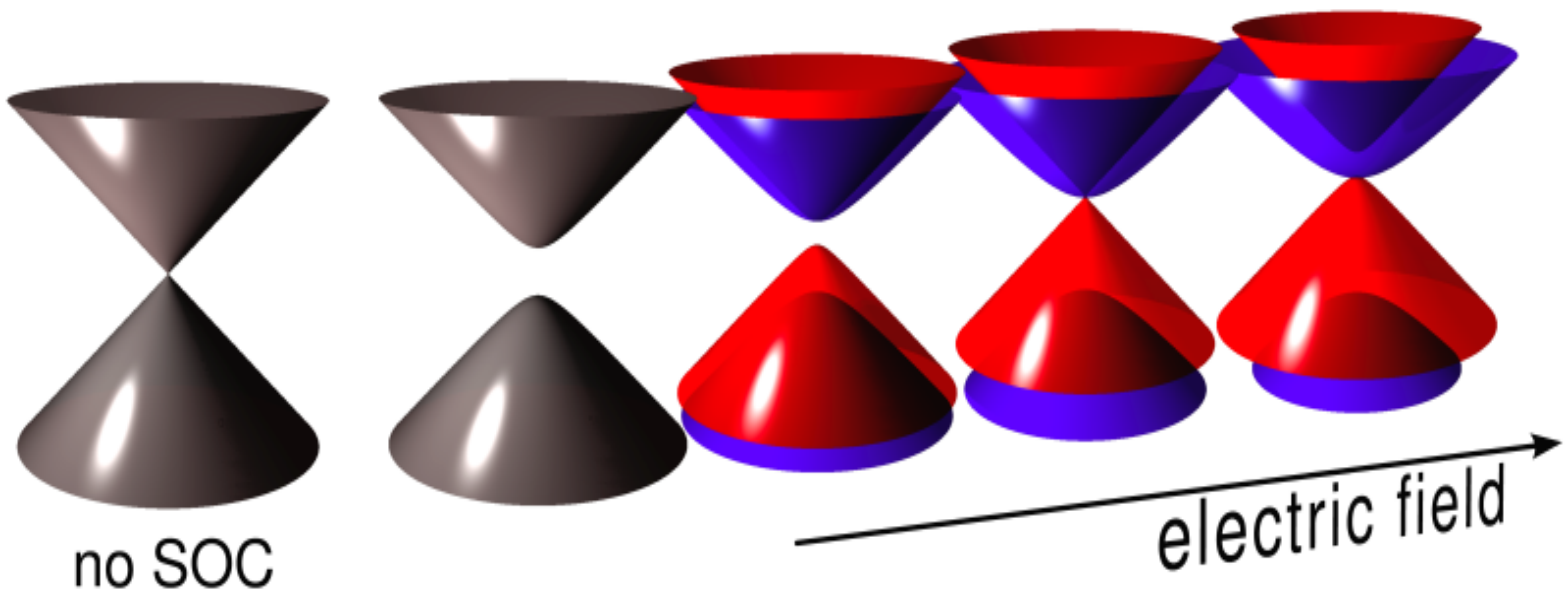

**Figure 3. Band structure topologies of graphene with spin-orbit coupling in a transverse electric field.**

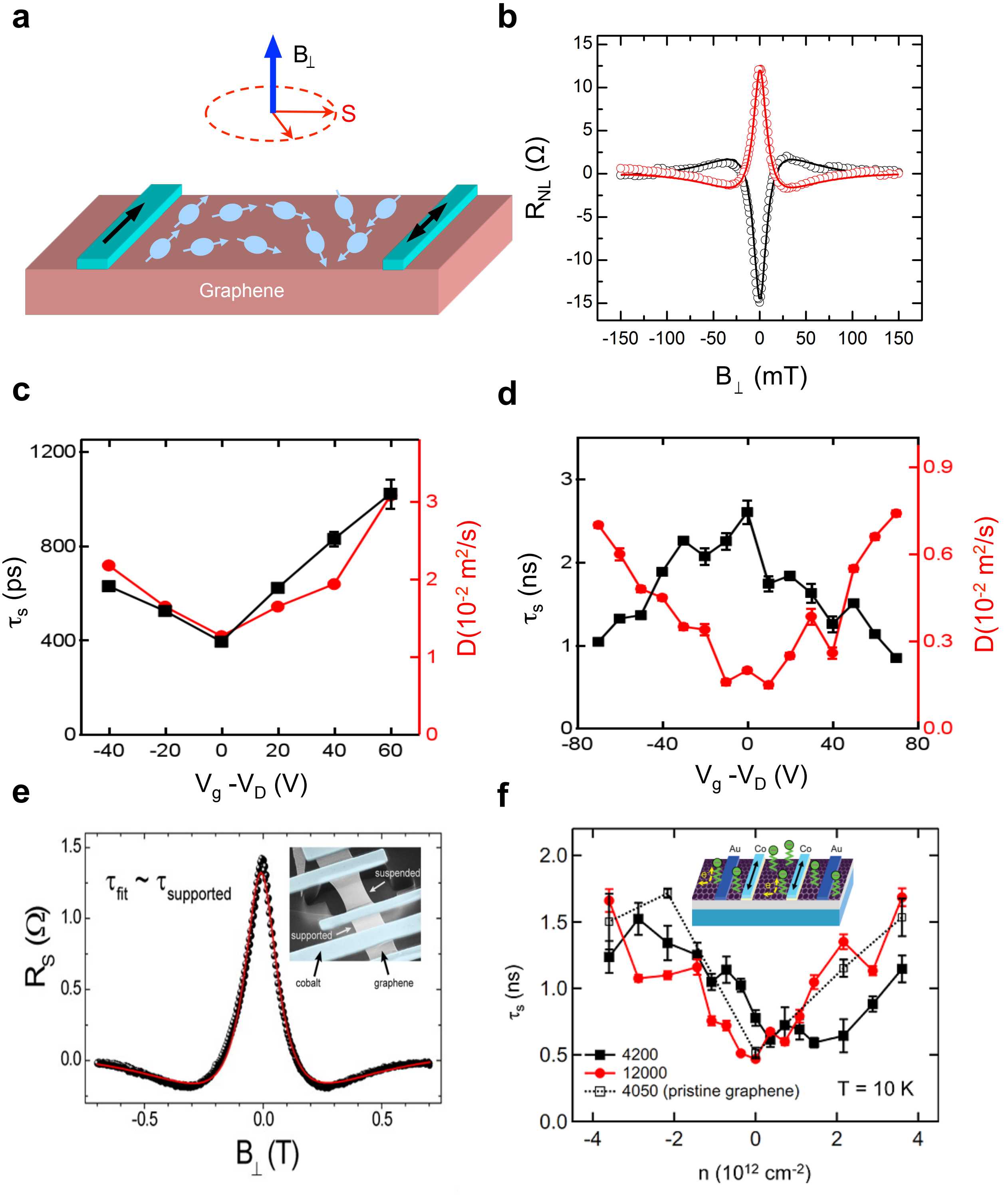

**Figure 4. Experimental studies of spin relaxation in graphene.**

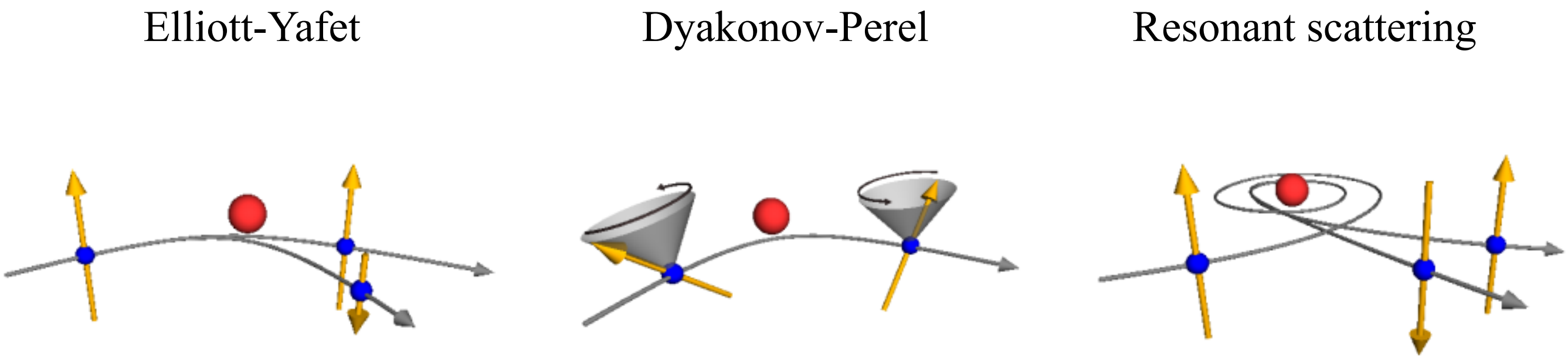

**Figure 5. Spin relaxation mechanisms in graphene.**

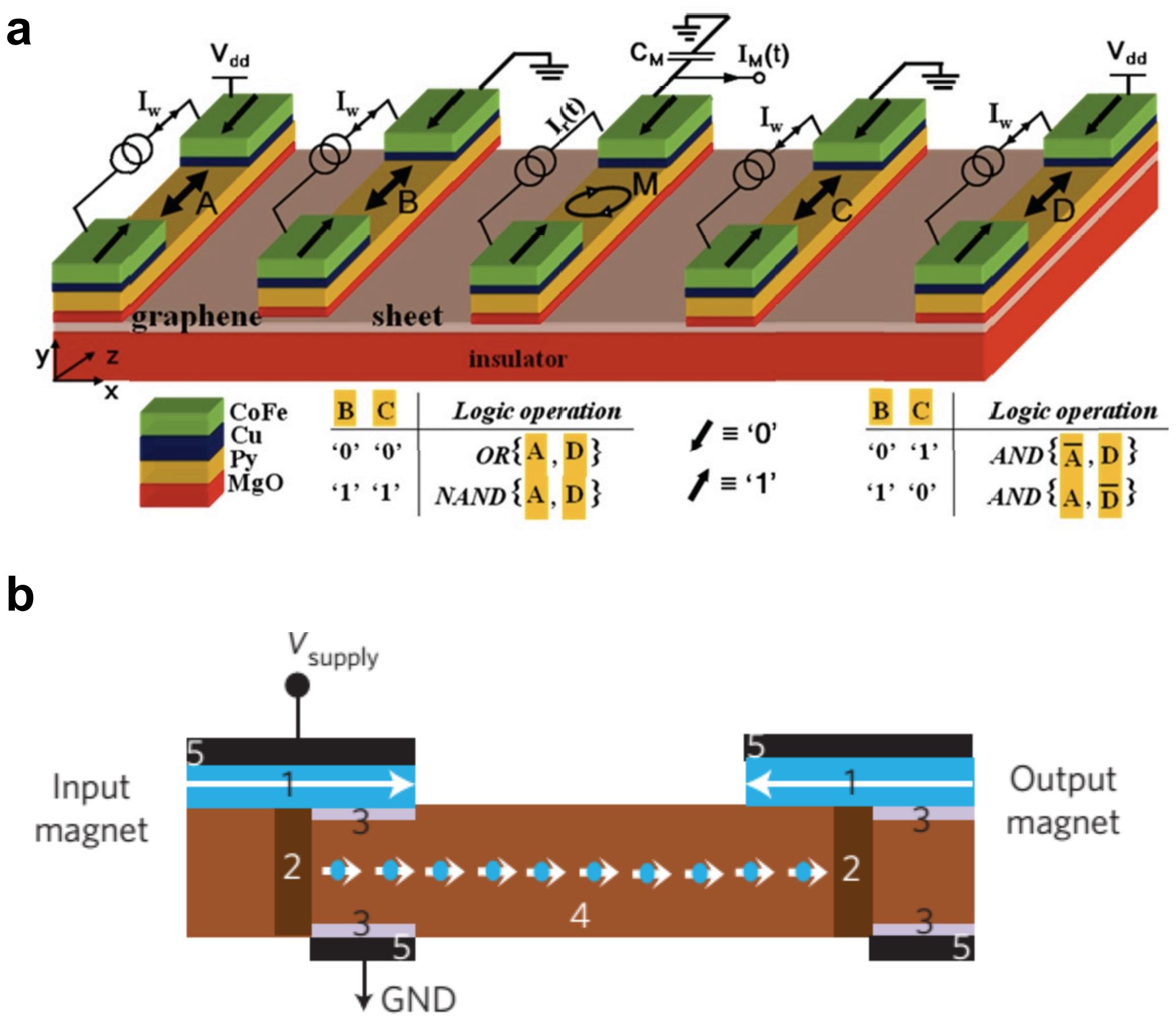

**Figure 6. Spin logic application of graphene spin valves.**